\documentclass[[aps,
prl,
reprint,
superscriptaddress,
groupedaddress,
amsmath,amssymb,
floatfix,
12pt, 
]{revtex4}
\usepackage{color,soul} 
\usepackage{graphicx}
\usepackage{wrapfig}
\usepackage{dcolumn}
\usepackage{bm}
\usepackage{seqsplit} 
\usepackage{ulem} 
\usepackage{float}

\begin{document}


\title{First-principles calculations of phonon transport across a vacuum gap}

\author{Takuro Tokunaga}
\affiliation{
Department of Mechanical Engineering, University of Utah, Salt Lake City, Utah 84112, United States.}
 
\author{Masao Arai}
\affiliation{
International Center for Materials Nanoarchitectonics (MANA), National Institute for Materials Science (NIMS), 1-1 Namiki, Tsukuba, Ibaraki 305-0044, Japan.}

\author{Kazuaki Kobayashi}
\affiliation{
International Center for Materials Nanoarchitectonics (MANA), National Institute for Materials Science (NIMS), 1-1 Namiki, Tsukuba, Ibaraki 305-0044, Japan.}

\author{Wataru Hayami}
\affiliation{
International Center for Materials Nanoarchitectonics (MANA), National Institute for Materials Science (NIMS), 1-1 Namiki, Tsukuba, Ibaraki 305-0044, Japan.}

\author{Shigeru Suehara}
\affiliation{
International Center for Materials Nanoarchitectonics (MANA), National Institute for Materials Science (NIMS), 1-1 Namiki, Tsukuba, Ibaraki 305-0044, Japan.}

\author{Takuma Shiga}
\email{shiga@photon.t.u-tokyo.ac.jp}
\affiliation{
Department of Mechanical Engineering, University of Tokyo, Bunkyo, Tokyo 113-8656, Japan.}

\author{Keunhan Park}
\email{kpark@mech.utah.edu}
\affiliation{
Department of Mechanical Engineering, University of Utah, Salt Lake City, Utah 84112, United States.}

\author{Mathieu Francoeur}
\email{mfrancoeur@mech.utah.edu}
\affiliation{
Department of Mechanical Engineering, University of Utah, Salt Lake City, Utah 84112, United States.}


\begin{abstract}
Phonon transport across a vacuum gap separating intrinsic silicon crystals is predicted via the atomistic Green's function method combined with first-principles calculations of all interatomic force constants. The overlap of electron wave functions in the vacuum gap generates weak covalent interaction between the silicon surfaces, thus creating a pathway for phonons. Phonon transport, dominated by acoustic modes, exceeds near-field radiation for vacuum gaps smaller than $\sim$1 nm. The first-principles-based approach proposed in this work is critical to accurately quantify the contribution of phonon transport to heat transfer in the extreme near field. 

\end{abstract}

\maketitle
\newpage

Near-field radiation heat transfer across a sub-wavelength vacuum gap can exceed Planck's far-field blackbody limit owing to tunneling of evanescent electromagnetic waves \cite{Polder1971}. Near-field radiation research has been primarily driven by potential applications in energy conversion \cite{DiMatteo2001,Fiorino2018_NatNanotech,Inoue2019,Bhatt2020,Lucchesi2021_NanoLett,Mittapally2021}, thermal management \cite{Otey2010,Ben-Abdallah2015,Ito2017,Elzouka2017,Fiorino2018_ACSNano,Tranchant2019}, and near-field thermal spectroscopy \cite{Wilde2006,Jones2012,Babuty2013}, amongst others. The framework of fluctuational electrodynamics \cite{Rytov1978}, in which the macroscopic Maxwell equations are supplemented by thermally fluctuating currents, generally well describes near-field radiation experiments down to nanosized vacuum gaps \cite{Shen2009,Rousseau2009,Ottens2011,Kim2015,St-Gelais2014,St-Gelais2016,Song2016,Bernardi2016,Ito2017,Fiorino2018_NanoLett,Lim2018,Ghashami2018,DeSutter2019,Shi2019,Lim2020,Sabbaghi2020,Salihoglu2020,Tang2020,Ying2020,Lucchesi2021}. However, a few experiments carried out at single-digit nanometer vacuum gaps have reported heat transfer largely exceeding fluctuational electrodynamics predictions \cite{Kloppstech2017,Cui2017,Messina2018,Jarzembski2019}, thus suggesting that not only electromagnetic waves can contribute to thermal transport in the extreme near field. 

Prior theoretical \cite{Prunnila2010,Budaev2011,Sellan2012,Xiong2014,Ezzahri2014,Chiloyan2015,Pendry2016,Sasihithlu2017,Alkurdi2020,Volokitin2021,Chen2021,Tokunaga2021,ChenNagayama2021} and experimental \cite{Altfeder2010,Jarzembski2019} works reported that phonon transport across vacuum gaps can dominate heat transfer in the extreme near field. Since this transport mechanism is mediated by lattice vibrations, an all-atom model is the most effective approach for incorporating the three-dimensional (3D) interatomic force interactions into phonon transport calculations. The scattering boundary method and the atomistic Green's function (AGF) framework are both all-atom models that enable calculating transport of coherent phonons inside materials \cite{Sadasivam2014,Li2019}, across interfaces \cite{Zhang2007,Zhang2007_3D,Landry2010,Tian2012,Latour2017}, and across vacuum gaps \cite{Sellan2012, Xiong2014, Ezzahri2014, Chiloyan2015, Alkurdi2020,Xiong2020,Tokunaga2021}.

Phonon transport across a vacuum gap separating intrinsic silicon (Si) \cite{Sellan2012,Alkurdi2020}, doped Si \cite{Ezzahri2014}, cubic polytype of silicon carbide \cite{Ezzahri2014}, and gold \cite{Alkurdi2020} bulk materials has been predicted via the scattering boundary method. In all cases, the interatomic force constants connecting the materials across the vacuum gap were calculated using empirical potential models. It was shown that phonon transport can generally exceed near-field radiation below vacuum gaps of $\sim$1 nm. Xiong {\it{et al.}} \cite{Xiong2014} applied for the first time the 3D AGF method for calculating extreme near-field heat transfer between two silica clusters. Using the van Beest, Kramer, and van Santen potential \cite{vanBeest1990} for modeling the interatomic force constants, it was found that for vacuum gaps $d$ below 0.4 nm, the conductance varied as $d^{-12}$ which cannot be explained with fluctuational electrodynamics. This regime was interpreted as phonon transport via pseudocovalent bonds formed by the overlap of electron wave functions in the vacuum gap. Chiloyan {\it{et al.}} \cite{Chiloyan2015} proposed a unified framework based on the 3D AGF method and the microscopic Maxwell's equations for predicting the transition from near-field radiation to conduction at contact between two sodium chloride crystals. The short-range repulsive forces in sodium chloride were modeled via the Kellerman potential \cite{Kellermann1940}, whereas the long-range Coulomb forces coupling the two crystals across the vacuum gap were obtained from the microscopic Maxwell's equations and the harmonic force constants of Jones and Fuchs \cite{Jones1971}. For a temperature of 55 K and vacuum gaps below 1 nm, a heat transfer coefficient exceeding fluctuational electrodynamics predictions was reported owing to the increasing contribution of acoustic phonon to heat transport. For vacuum gaps between 1 and 2.8 nm, the heat transfer coefficient followed a $d^{-2}$ power law which is the signature of surface phonon-polaritons, defined as the coupling of transverse optical phonons and evanescent electromagnetic waves.

To date, all-atom models applied for predicting phonon transport across vacuum gaps rely on empirical potential models for calculating the interatomic force constants \cite{Sellan2012,Xiong2014,Ezzahri2014,Chiloyan2015,Alkurdi2020,Tokunaga2021}. With such an approach, however, the interatomic force constants, and therefore the heat transfer, are highly-dependent on empirical parameters. In addition, the definition of a cutoff vacuum gap beyond which the materials are not connected anymore by interatomic forces hinders the physics behind the transition from near-field radiation to phonon transport \cite{Sellan2012}. 

The objective of this work is therefore to predict phonon transport across vacuum gaps via the 3D AGF method combined with first-principles calculations of the interatomic force constants. In that way, it will be possible to accurately predict the vacuum gap below which phonon transport exceeds near-field radiation. Specifically, the case of two intrinsic Si crystals separated by a vacuum gap is considered, where all interatomic force constants in Si and in the vacuum gap are obtained from first-principles calculations. The results reveal that phonon transport exceeds near-field radiation at room temperature for vacuum gaps approximately equal to or smaller than  1 nm. It is shown that phonon transport is primarily mediated by acoustic phonons across weak covalent interaction formed in the vacuum gap by overlapping electron wave functions.

\begin{figure}[p!]
\centering
\includegraphics[width=0.9\linewidth]{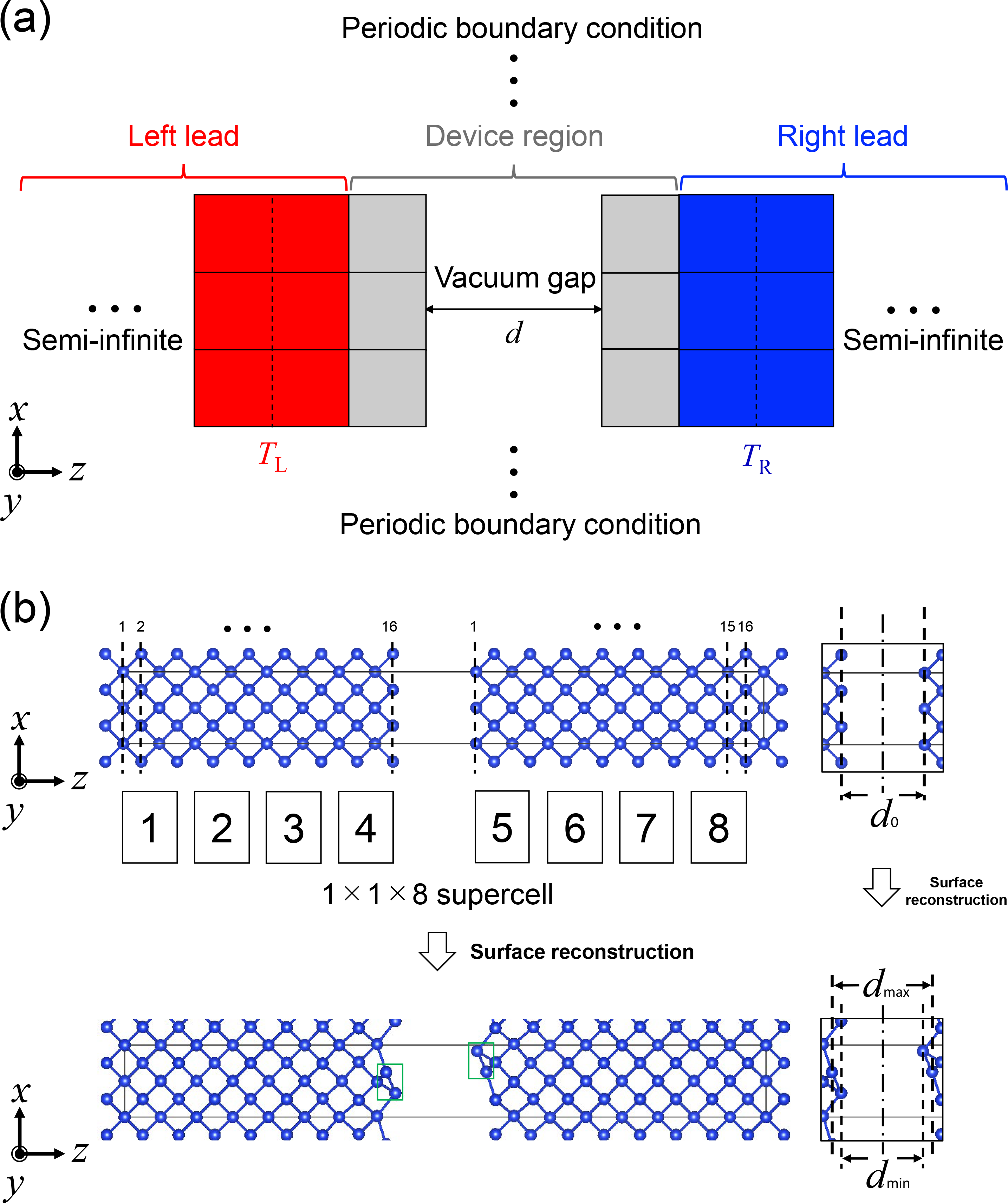}
\caption{(a) Schematic of the system used for calculating phonon heat transfer across the device region connected to the semi-infinite left and right leads maintained at temperatures $T_\mathrm{L}$ = 305 K and $T_\mathrm{R}$ = 300 K. (b) A 1 $\times$ 1 $\times$ 8 supercell, which is the combination of eight conventional unit cells (black numbered boxes), is used for calculating the interatomic force constants of the device region. Surface reconstruction of p(2 {$\times$} 1) asymmetric dimer (green boxes) is formed after structural optimization.}
\label{Fig.1}
\end{figure} 

Figure \ref{Fig.1}(a) shows the system used for calculating the phonon heat transfer coefficient, where intrinsic Si of (001) orientation is placed along the $z$-direction. The system is composed of three parts: the semi-infinite left and right leads maintained at constant and uniform temperatures of $T_{\mathrm{L}}$ = 305 K and $T_{\mathrm{R}}$ = 300 K, and the device region that includes the vacuum gap of thickness $d$. Periodic boundary conditions are applied to the transverse $x$- and $y$-directions \cite{Tian2012}. The heat transfer coefficient due to phonon transport across the vacuum gap along the $z$-direction is calculated via the 3D AGF method as follows \cite{Zhang2007_3D, Latour2017}: 
\begin{equation}
\label{phonon heat flux}
h_\mathrm{ph}  \ = \ \frac{1}{A(T_\mathrm{L}-T_\mathrm{R})}\int_{0}^{\infty}d{\omega}\frac{\hbar\omega}{2\pi}{\cal T_\mathrm{ph}}({\omega})[N(\omega, T_\mathrm{L})-N(\omega,T_\mathrm{R})]
\end{equation} 
where $A$ is the cross-sectional area of the unit cell, $\omega$ is the phonon frequency, and $N\left(\omega, T_{j}\right) = 1/[\mathrm{exp}({\hbar\omega/k_\mathrm{B}T_{j}})-1]$ $\left(j = \mathrm{L, R}\right)$ is the Bose-Einstein distribution function in which $\hbar$ and $k_{\mathrm{B}}$ are respectively the reduced Planck constant and the Boltzmann constant. The phonon transmission function, $\cal T_\mathrm{ph}({\omega})$, is given by $\left(1/N_{\textbf{{k}}_{||}}\right){\cdot}\sum_{\textbf{{k}}_{||}}\widetilde{\cal T}_\mathrm{ph}({\omega},{\textbf{{k}}_{||}})$, where $\textbf{{k}}_{||}$ is the wave vector parallel to the $x-y$ plane, and $N_{\textbf{{k}}_{||}}$ is the number of discrete parallel wave vectors within the first Brillouin zone. The phonon transmission function per unit parallel wave vector, $\widetilde{\cal T}_\mathrm{ph}({\omega},{\textbf{{k}}_{||}})$, is given by the Caroli formula \cite{Caroli1971}:
\begin{equation}
\label{transmission function}
{\widetilde{\cal T}_\mathrm{ph}}({\omega},\textbf{{k}}_{||}) \ = \ \mathrm{Trace}[\Gamma_\mathrm{L}G_\mathrm{d}\Gamma_\mathrm{R}G_\mathrm{d}^\mathrm{\dagger}]
\end{equation} 
where the superscript $\dagger$ denotes conjugate transpose, $\Gamma_{j}({\omega},\textbf{{k}}_{||})$ ($j = \mathrm{L}, \mathrm{R}$) is the escape rate of phonons from the device region to the semi-infinite leads, and $G_\mathrm{d}({\omega},\textbf{{k}}_{||})$ is the Green's function of the device region. The interatomic force constants of the vacuum region and those due to the interactions with the left/right leads are incorporated into $G_\mathrm{d}({\omega},\textbf{{k}}_{||})$ \cite{Zhang2007_3D, Latour2017}.

First-principles calculations of all interatomic force constants are performed with the open-source density functional theory (DFT) package ABINIT \cite{Gonze2002, Gonze2020}. Prior works \cite{Sasihithlu2017,Alkurdi2020} have reported that phonon transport across vacuum gaps is dominated by van der Waals (vdW) interaction. As such, the impact of vdW interaction on the phonon heat transfer coefficient is also investigated hereafter. The DFT calculations are conducted under the generalized gradient approximation of Perdew-Burke-Ernzerhof \cite{PerdewBurkeErnzerhof1996} for the exchange and correlation functional with or without the vdW interaction correction by the semi-empirical DFT-D2 method \cite{Grimme2006}. The norm-conserving pseudopotential is used. The plane-wave energy cutoff for the norm-conserving pseudopotential is set to 10 Hartree (Ha), and 4~${\times}$~4~${\times}$~1 Monkhorst-Pack $k$-point mesh is used. The lattice constant is defined as 0.547 nm, thus minimizing the ground state total energy. The Gaussian smearing method is employed with a broadening width of 0.01 Ha. The convergence criterion of the structural optimization is set to less than $\mathrm{1~\times~10^{-6}}$ Ha/Bohr.

The interatomic force constants across the vacuum gap in the device region are calculated via the 1~${\times}$~1~${\times}$~8 supercell shown in Fig.~\ref{Fig.1}(b). The interatomic force constants of cells 4 and 5 from the 1~${\times}$~1~${\times}$~8 supercell are extracted after structural optimization forming a surface reconstruction of p(2 {$\times$} 1) asymmetric dimer \cite{Ramstad1995,Sagisaka2005}. A total of 16 layers on the left and right-hand sides of the 1~${\times}$~1~${\times}$~8 supercell are sufficient for eliminating the interactions with adjacent periodic cells along the $z$-direction \cite{Yang2005, Leahy2006}. In the DFT calculations, an atomic layer is defined as a $x-y$ plane containing Si atoms with a thickness equals to the diameter of a single atom.

The vacuum gap is the distance along the $z$-direction between the vertical cross-sections of the atoms located in the left and right leads that are adjacent to vacuum: see Fig.~\ref{Fig.1}(b). For initial vacuum gaps $d_{\mathrm{0}}$ of 0.30, 0.50, 0.70, 0.90, and 1.0 nm, the nearest ($d_{\mathrm{min}}$) and farthest ($d_{\mathrm{max}}$) distances after structural optimization respectively become: $\left(d_{\mathrm{min}}, d_{\mathrm{max}}\right)$ = $\mathrm{\left(0.40{\;}nm, 0.54{\;}nm\right)}$, $\mathrm{\left(0.62{\;}nm, 0.76{\;}nm\right)}$, $\mathrm{\left(0.82{\;}nm, 0.96{\;}nm\right)}$, $\mathrm{\left(1.02{\;}nm, 1.16{\;}nm\right)}$, and $\mathrm{\left(1.12{\;}nm, 1.26{\;}nm\right)}$. The vacuum gap $d$, defined as the average of $d_{\mathrm{min}}$ and $d_{\mathrm{max}}$, is used hereafter for analyzing the results ($d$= 0.47, 0.69, 0.89, 1.09, and 1.19 nm).

\begin{figure}[p!]
\centering
\includegraphics[width=1\linewidth]{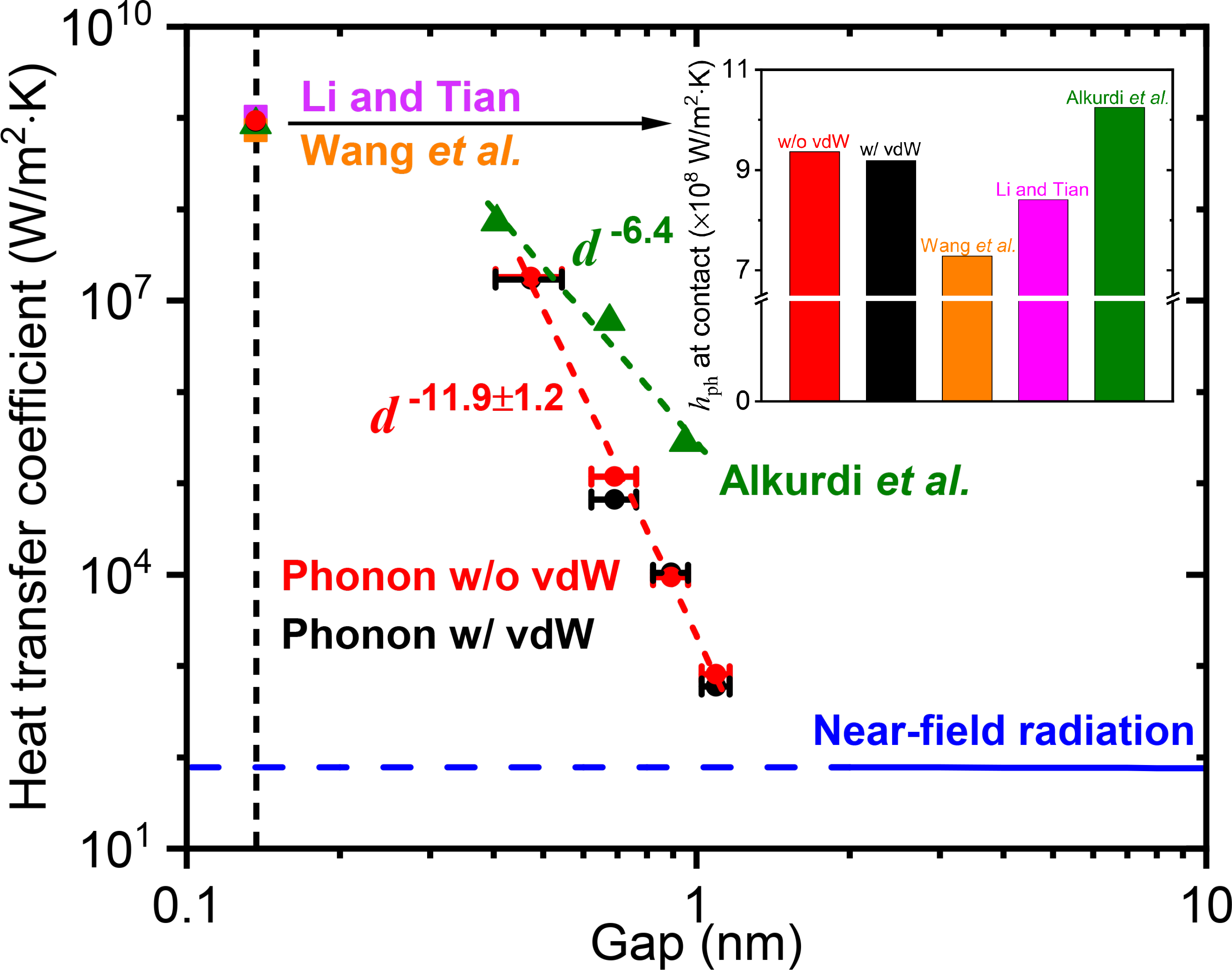}
\caption{Phonon heat transfer coefficient between intrinsic Si crystals in contact and separated by vacuum gaps $d$ of 0.47, 0.69, 0.89, and 1.09 nm with (w/) and without (w/o) vdW interaction. The error bars along the horizontal axis are derived from the best fits for $d_{\mathrm{min}}$ and $d_{\mathrm{max}}$. The results are compared against phonon heat transfer coefficients from the literature \cite{Wang2007, Li2019, Alkurdi2020}, and near-field radiation predictions based on fluctuational electrodynamics. Below a vacuum gap of 2 nm, the heat transfer coefficient due to near-field radiation is shown via a dashed line to emphasize the possible inaccuracy of the predictions made with a local dielectric function. The inset shows the phonon heat transfer coefficient, $h_{\mathrm{ph}}$, when the Si surfaces are in contact.}
\label{Fig.2}
\end{figure} 

Figure \ref{Fig.2} shows the phonon heat transfer coefficient, with and without vdW interaction, between Si crystals in contact and separated by a vacuum gap $d$. The extraction of the interatomic force constants for the case where the Si surfaces are in contact is described in Sec. S1 of the Supplemental Material \cite{Supplemental_Material}. The gap distance at contact, defined as the inter-planar spacing in the (001) orientation \cite{Chiloyan2015,Xiao2017}, is 0.137 nm. Note that surface reconstruction is not considered for the contact case. The heat transfer coefficient due to near-field radiation predicted by fluctuational electrodynamics is also plotted in Fig. \ref{Fig.2} (see Sec. S2 of the Supplemental Material for near-field radiation modeling \cite{Supplemental_Material}). 

The phonon heat transfer coefficients with and without vdW interaction for conduction at contact are respectively 9.19$\times10^{8}$ and 9.36$\times10^{8}$ $\mathrm{W/m^{2}{\cdot}K}$. This is in good agreement with prior works \cite{Wang2007,Li2019,Alkurdi2020} despite the different descriptions of the interatomic force interactions: see the inset of Fig. \ref{Fig.2}. At contact, phonon transport is fully ballistic through the device region and is determined by the phonon dispersion relation of crystalline bulk Si. Even if the maximum frequencies of crystalline Si are slightly different for first-principles and empirical-based calculations, the overall features of the phonon dispersion relation are similar, thus resulting in the agreement observed in Fig. \ref{Fig.2}. 

When the Si surfaces are separated by a vacuum gap, the phonon heat transfer coefficient with and without vdW interaction is approximately one order of magnitude larger than near-field radiation for $d$ = 1.09 nm, and increases monotonically as $d$ decreases. The phonon heat transfer coefficient vanishes for the largest vacuum gap considered ($d$ = 1.19 nm). Here, the near-field radiation heat transfer coefficient saturates as $d$ decreases since intrinsic Si does not support surface polaritons. Note that the near-field radiation result is plotted with a dashed line below a vacuum gap of 2 nm because the accuracy of fluctuational electrodynamics with a local dielectric function is questionable in this range. In the vacuum gap range of from 0.47 to 1.09 nm, the phonon heat transfer coefficient without vdW interaction follows a $d^{-11.9 \pm 1.2}$ power law. In contrast, by modeling the interatomic force constants in the vacuum gap via the Lennard-Jones potential, a power law of $d^{-6.4}$ is retrieved from the data of Alkurdi {\it{et al.}} \cite{Alkurdi2020} for separation distances of from $\sim$0.41 nm to $\sim$0.95 nm. Here, however, the predicted phonon heat transfer coefficient is nearly independent of the contribution of the vdW interaction. The vdW interaction is not strong enough to connect the two Si surfaces at vacuum gap distances larger than $\sim$1 nm. For distances smaller than $\sim$1 nm, it is hypothesized that the contribution of the vdW interaction is negligible compared to the weak covalent interaction induced by the overlap of electron wave functions in the vacuum gap. This hypothesis is verified hereafter by analyzing the spatial distribution of electron density.

\begin{figure}[p!]
\centering
\includegraphics[width=0.9\linewidth]{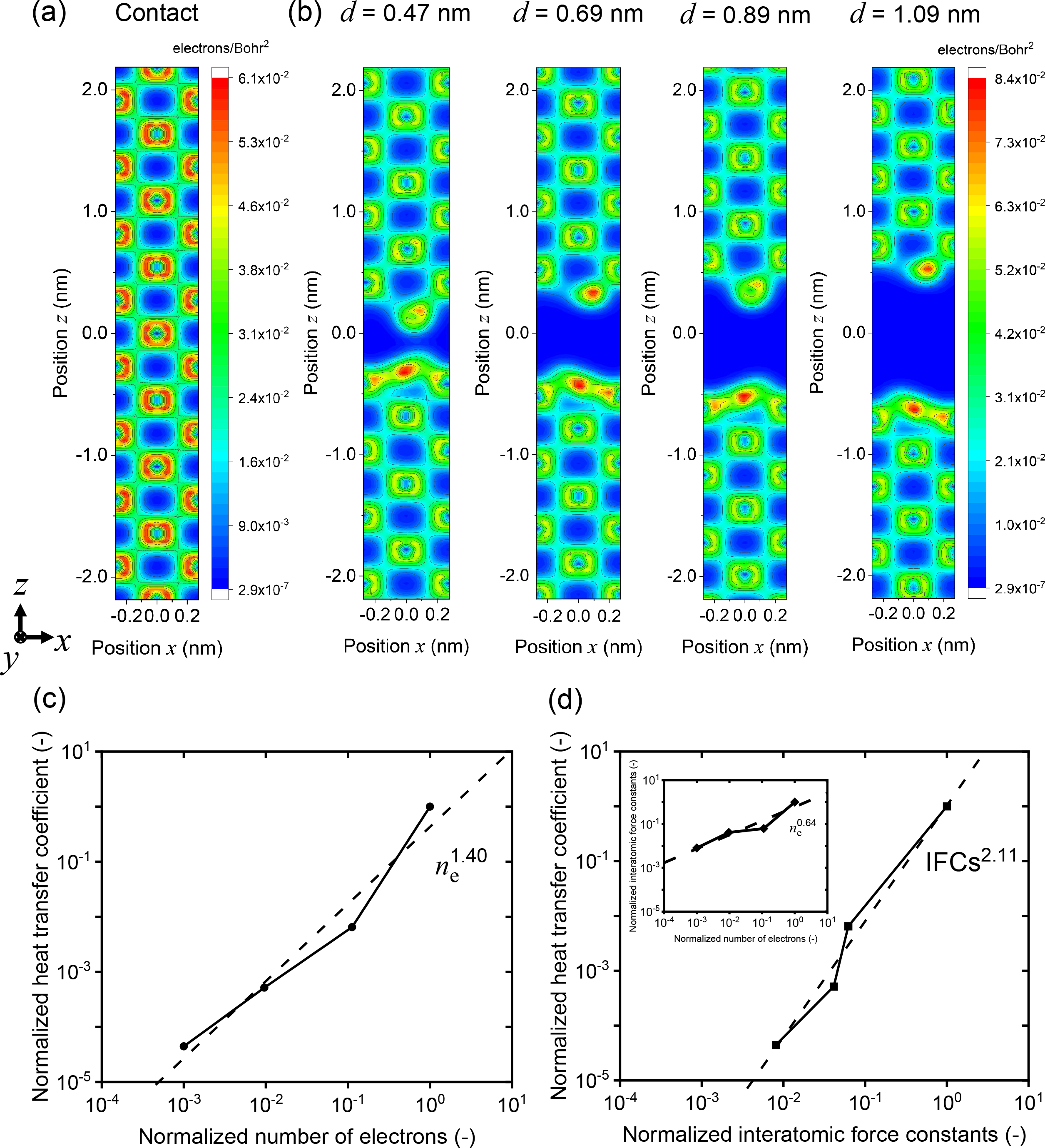}
\caption{(a) Spatial distribution of electron density within the supercell for the device region when the Si surfaces are in contact, and (b) for vacuum gaps $d$ of 0.47, 0.69, 0.89, and 1.09 nm. The position $z$ = 0 corresponds to the middle of the vacuum gap region, and 1 Bohr = 0.0529 nm. (c) Phonon heat transfer coefficient as a function of the number of electrons $n_\mathrm{e}$ in the middle of the vacuum gap. (d) Phonon heat transfer coefficient as a function of the interatomic force constants (IFCs), and interatomic force constants as a function of the number of electrons in the middle of the vacuum gap (inset). The heat transfer coefficient, interatomic force constants, and number of electrons are normalized by their respective maximum at $d$ = 0.47 nm.}
\label{Fig.3}
\end{figure} 

The spatial distribution of electron density within the supercell for the device region is shown in Fig. \ref{Fig.3}(a) at contact, and in Fig. \ref{Fig.3}(b) for vacuum gaps of 0.47, 0.69, 0.89, and 1.09 nm. The electron density is periodically distributed when the Si surfaces are in contact, whereas non-symmetric distributions are observed when there is a vacuum gap. This is due to the different spatial distribution of Si atoms in the supercell arising from surface reconstruction. Physically, a non-zero electron density in the vacuum gap implies that the two Si surfaces are connected through weak covalent interaction. Here, Si atoms of p(2 {$\times$} 1) reconstruction form $\pi$ bonds for stabilizing the dimer \cite{Bolding1990,Ramstad1995}. Note that non-contact imaging for atomic force microscopy of Si(100) surface of p(2 {$\times$} 1) structure experimentally detected surface interaction up to vacuum gaps of $\sim$1 nm by measuring force gradient \cite{Kitamura1996}. Further numerical simulations suggested that the force gradient was dominated by weak covalent interaction \cite{Perez1997, Perez1998}.

Figure \ref{Fig.3}(c) shows the phonon heat transfer coefficient as a function of the number of electrons, $n_\mathrm{e}$, in the middle of the vacuum gap (Sec. S3 of the Supplemental Material \cite{Supplemental_Material} describes how $n_\mathrm{e}$ is calculated). Both quantities are normalized by their respective maximum obtained at $d$ = 0.47 nm. The heat transfer coefficient increases as the number of electrons increases, following a $n_\mathrm{e}^{1.40}$ power law. Figure \ref{Fig.3}(d) shows that the phonon heat transfer coefficient increases proportionally to approximately the square of the interatomic force constants, whereas the interatomic force constants increase as $n_\mathrm{e}^{0.64}$ (Sec. S3 of the Supplemental Material \cite{Supplemental_Material} describes the quantification of the interatomic force constants). Combining these two results reveal that the phonon heat transfer coefficient increases as $n_\mathrm{e}^{1.35}$. This is essentially the same relationship as the one observed in Fig. \ref{Fig.3}(c), confirming that the interatomic force constants driving phonon transport are mediated by the overlapping electron wave functions in the vacuum gap. Physically, phonon transport occurs through weak covalent interaction formed in the vacuum gap connecting the Si crystals.  

\begin{figure}[p!]
\centering
\includegraphics[width=0.9\linewidth]{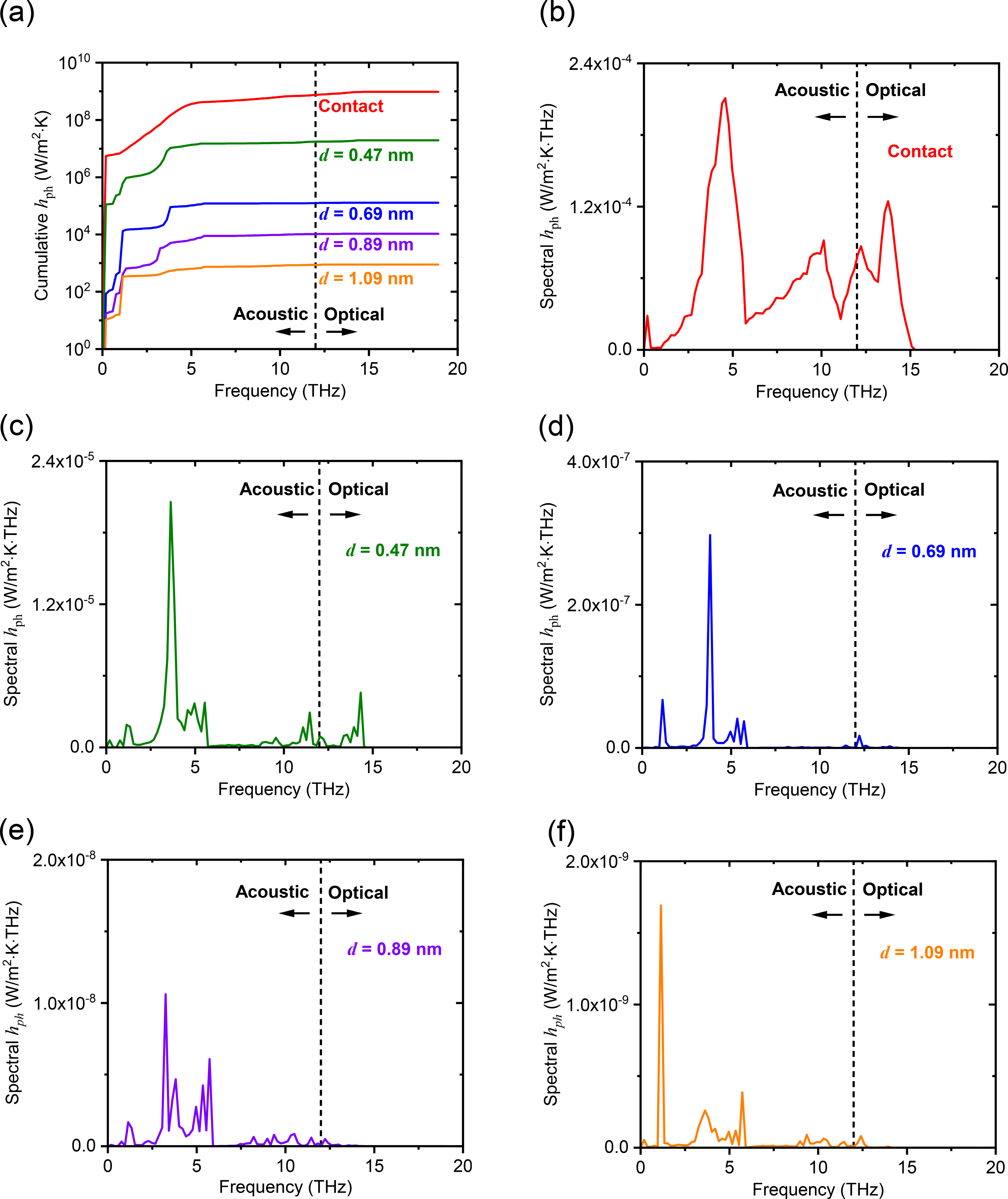}
\caption{
(a) Cumulative phonon heat transfer coefficient as a function of frequency at contact and for vacuum gaps $d$ of 0.47, 0.69, 0.89, and 1.09 nm. The corresponding spectral phonon heat transfer coefficients are shown in panels (b) to (f).}
\label{Fig.4}
\end{figure} 

The cumulative phonon heat transfer coefficient as a function of frequency is presented in Fig.~\ref{Fig.4}(a) at contact and for vacuum gaps of 0.47, 0.69, 0.89, and 1.09 nm, whereas the corresponding spectral phonon heat transfer coefficients are shown in Figs. \ref{Fig.4}(b)--(f). The vertical dashed line at 12 THz delimits the frequencies associated with acoustic and optical phonons, as determined by the phonon dispersion relations of bulk Si predicted via DFT calculations (see Sec. S4 of the Supplemental Material \cite{Supplemental_Material}). Acoustic phonons (frequencies smaller than 12 THz) dominate the heat transfer coefficient for all cases. At contact, 76.7{\%} of the heat transfer coefficient is mediated by acoustic phonons, and this proportion increases to more than 88.3{\%} when there is a vacuum gap. This is explained by the larger acoustic phonon population at room temperature, calculated as the product of the phonon density of states and Bose-Einstein distribution, with respect to the optical phonon population (see Sec. 5 of the Supplemental Material \cite{Supplemental_Material}). This demonstrates that heat transport between Si surfaces separated by vacuum gaps approximately equal to or smaller than 1 nm is essentially mediated by acoustic phonons across weak covalent interaction formed in the vacuum gap. 

In summary, phonon transport across a vacuum gap separating two intrinsic Si crystals has been predicted via the 3D AGF method and first-principles calculations, based on the density functional theory, of all interatomic force constants. For a vacuum of 1.09 nm, the phonon heat transfer coefficient, dominated by acoustic modes, exceeds near-field radiation predictions based on fluctuational electrodynamics by approximately one order of magnitude, and increases as the vacuum gap decreases at a rate of $d^{-11.9 \pm 1.2}$. The overlapping electron wave functions form weak covalent interaction connecting the two Si surfaces, thus inducing phonon transport across vacuum gaps. First-principles calculations of interatomic force constants enable incorporating surface electronic states in a precise manner into phonon transmission calculations. The extension of the framework presented in this work to metallic surfaces could provide further interpretations of previous extreme near-field heat transfer experiments \cite{Kloppstech2017,Cui2017,Messina2018}. 

\begin{acknowledgments}
This work was supported by the National Science Foundation (Grants No. CBET-1605584 and No. CBET-1952210). T.T. also appreciates the financial support by the Yamada Science Foundation and the Fujikura Foundation. We thank Prof. Akiko Kaneko, Prof. Takeo Fujiwara, Prof. Takuma Hori, and Dr. Nobuo Tajima for fruitful discussions. K.P. gratefully acknowledges support from the National Research Foundation of Korea (NRF) grants funded by the Korea government (MSIT) (No. 2020R1A3B2079741; 2020R1A4A3079853) for his research at Pohang University of Science and Technology (POSTECH) as a visiting professor. The DFT calculations in this study were mainly performed on Numerical Materials Simulator at the National Institute for Materials Science (NIMS) in Japan. The resources from the Center for High Performance Computing at the University of Utah and the support of Prof. Anita M. Orendt, Dr. Martin Cuma, and Dr. Wim R. Cardoen is also acknowledged. The three-dimensional schematics of crystal structures were generated using the open-source visualization software VESTA \cite{Momma2011}.  
\end{acknowledgments}

\normalem 
\bibliography{Ref}
\end{document}



\title{Supplemental Material \\First-principles calculations of phonon transport across a vacuum gap} 
\author{Takuro Tokunaga}
\affiliation{
Department of Mechanical Engineering, University of Utah, Salt Lake City, Utah 84112, United States.}
 
\author{Masao Arai}
\affiliation{
International Center for Materials Nanoarchitectonics (MANA), National Institute for Materials Science (NIMS), 1-1 Namiki, Tsukuba, Ibaraki 305-0044, Japan.}

\author{Kazuaki Kobayashi}
\affiliation{
International Center for Materials Nanoarchitectonics (MANA), National Institute for Materials Science (NIMS), 1-1 Namiki, Tsukuba, Ibaraki 305-0044, Japan.}

\author{Wataru Hayami}
\affiliation{
International Center for Materials Nanoarchitectonics (MANA), National Institute for Materials Science (NIMS), 1-1 Namiki, Tsukuba, Ibaraki 305-0044, Japan.}

\author{Shigeru Suehara}
\affiliation{
International Center for Materials Nanoarchitectonics (MANA), National Institute for Materials Science (NIMS), 1-1 Namiki, Tsukuba, Ibaraki 305-0044, Japan.}

\author{Takuma Shiga}
\email{shiga@photon.t.u-tokyo.ac.jp}
\affiliation{
Department of Mechanical Engineering, The University of Tokyo, Bunkyo, Tokyo 113-8656, Japan.}

\author{Keunhan Park}
\email{kpark@mech.utah.edu}
\affiliation{
Department of Mechanical Engineering, University of Utah, Salt Lake City, Utah 84112, United States.}

\author{Mathieu Francoeur}
\email{mfrancoeur@mech.utah.edu}
\affiliation{
Department of Mechanical Engineering, University of Utah, Salt Lake City, Utah 84112, United States.}



\maketitle
\newpage
\section{S1. First-principles calculations of phonon transport across a silicon interface}
\begin{figure}[p!]
\centering
\includegraphics[width=1\linewidth]{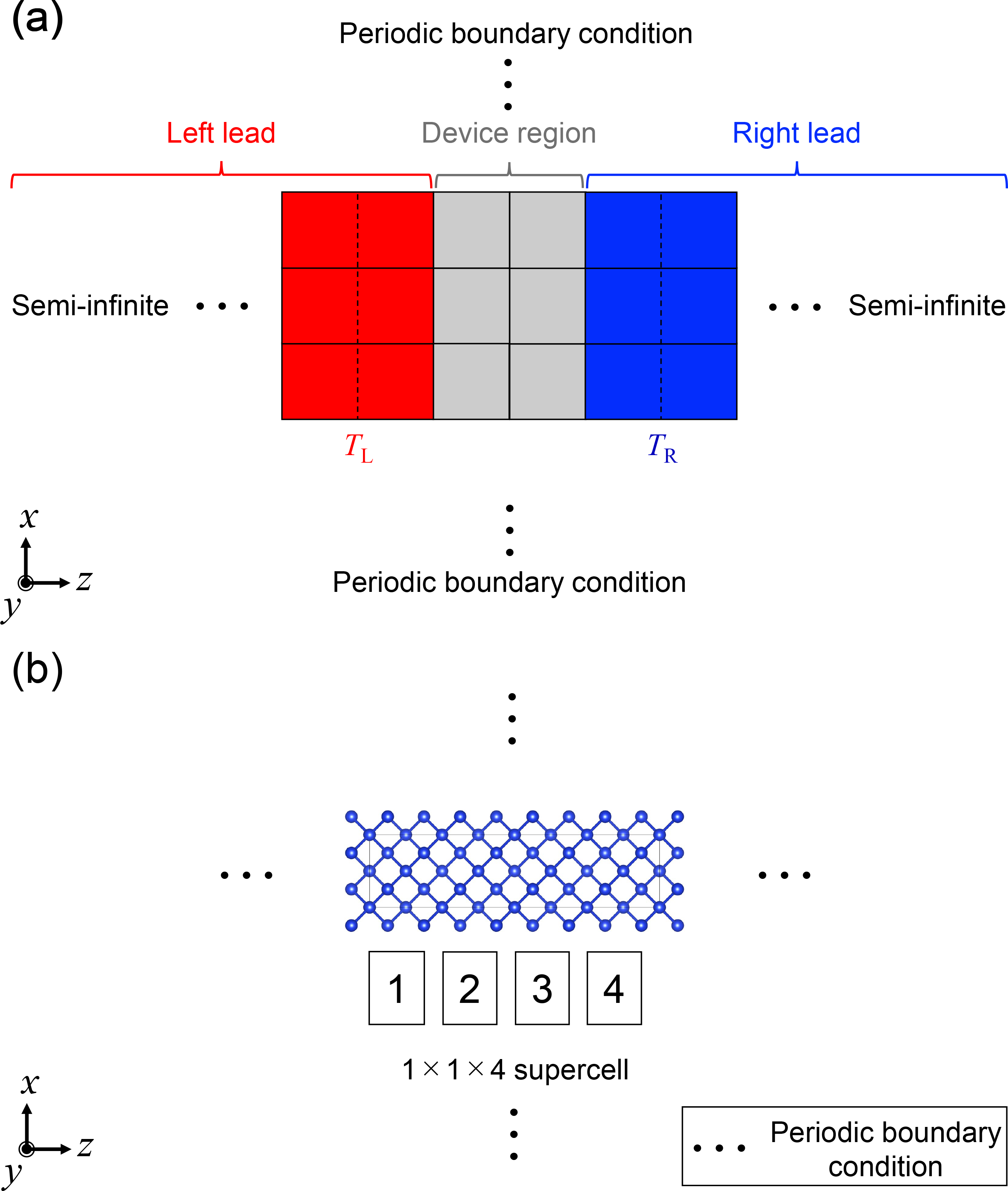}
\caption{(a) Schematic of the system used for calculating phonon heat transfer across an interface. The semi-infinite left and right leads are respectively maintained at temperatures $T_{\mathrm{L}}$ = 305 K and $T_{\mathrm{R}}$ = 300 K. (b) A 1~${\times}$~1~${\times}$~4 supercell, which is the combination of four conventional unit cells (black numbered boxes), is used for calculating the interatomic force constants.}
\label{Fig.S1}
\end{figure}

The system used for calculating phonon transport across an intrinsic silicon (Si) interface is shown in Fig.~S\ref{Fig.S1}(a). The system is the same as Fig. 1(a) of the main text, except that there is no vacuum gap region.

A 1~${\times}$~1~${\times}$~4 supercell shown in Fig.~S\ref{Fig.S1}(b) is used for computing the interatomic force constants via first-principles calculations based on the density functional theory. The interatomic force constants of cells 1 and 2 are extracted for the left and right leads, and also for the device region. The gap distance in contact is defined as the inter-planar spacing in the (001) orientation \cite{Chiloyan2015,Xiao2017} (i.e., 0.137 nm). Surface reconstruction is not considered when the Si crystals are in contact.

 \clearpage
\newpage

\section{S2. Near-field radiation modeling}
The heat transfer coefficient due to near-field radiation between the left (L) and right (R) semi-infinite leads separated by a vacuum gap $d$ is calculated using fluctuational electrodynamics \cite{Rytov1978,Polder1971}: 
\begin{equation}
\label{radiative flux}
h_\mathrm{rad}  \ = \ \frac{1}{\pi^{\mathrm{2}}(T_\mathrm{L}-T_\mathrm{R})}\int_{0}^{\infty}d{\omega}\left[	\Theta\left(\omega,T_\mathrm{L}\right) - \Theta\left(\omega,T_\mathrm{R}\right)\right]
\\ \\ 
\int_{0}^{\infty}dk_\mathrm{\rho}k_\mathrm{\rho}\sum_{\gamma=\mathrm{TE,TM}}{\cal T_\mathrm{rad}^{\gamma}}\left(\omega,k_\mathrm{\rho}\right)
\end{equation} 
where $T_{j}$ is the temperature of semi-infinite lead $j$ ($j = \mathrm{L}, \mathrm{R}$), $\omega$ is the angular frequency, $k_\mathrm{\rho}$ is the component of the wave vector parallel to a surface, and $\Theta(\omega, T_{j})$ is the mean energy of an electromagnetic state calculated as $\hbar\omega/[\mathrm{exp}({\hbar\omega/k_\mathrm{B}T_{j}})-1]$. The transmission functions in polarization state $\gamma$ for propagating ($k_\mathrm{\rho}$ $<$ $k_\mathrm{0}$) and evanescent ($k_\mathrm{\rho}$ $>$ $k_\mathrm{0}$) electromagnetic waves in vacuum are calculated as:
\begin{equation}
\label{propagating transmission}
{\cal T_\mathrm{rad,prop}^\gamma}\left(\omega,k_\mathrm{\rho}\right) \ = \frac{\left(1-\left|r_\mathrm{0L}^{\gamma}\right|^{2}\right)\left(1-\left|r_\mathrm{0R}^{\gamma}\right|^{2}\right)}{4\left|1-r_\mathrm{0L}^{\gamma}r_\mathrm{0R}^{\gamma}e^{2i\mathrm{Re}\left(k_{\mathrm{z0}}\right)d}\right|^{2}}
\end{equation}
\begin{equation}
\label{evanescent transmission}
{\cal T_\mathrm{rad,evan}^\gamma}\left(\omega,k_\mathrm{\rho}\right) \ = e^{-2\mathrm{Im}\left(k_\mathrm{z0}\right)d}\frac{\mathrm{Im}\left(r_\mathrm{0L}^{\gamma}\right)\mathrm{Im\left(r_\mathrm{0R}^{\gamma}\right)}}{\left|1-r_\mathrm{0L}^{\gamma}r_\mathrm{0R}^{\gamma}e^{-2\mathrm{Im}\left(k_{\mathrm{z0}}\right)d}\right|^{2}}
\end{equation} where $k_\mathrm{0}={\omega}/c_{0}$ is the magnitude of the vacuum wave vector with $c_{0}$ as the speed of light in vacuum, $k_\mathrm{z0}$ is the component of the vacuum wave vector perpendicular to an interface, and $r_{0j}^\gamma$ is the Fresnel reflection coefficient at the vacuum-lead $j$ ($j = \mathrm{L}, \mathrm{R}$) interface in polarization state $\gamma$ \cite{Yeh1988}. The local, frequency-dependent dielectric function of intrinsic Si provided in Refs. \cite{Adachi1988,Aoki1991} is used in the calculations.
\newpage

\section{S3. Number of electrons and interatomic force constants in the vacuum gap region}
\begin{figure}[p!]
\centering
\includegraphics[width=1\linewidth]{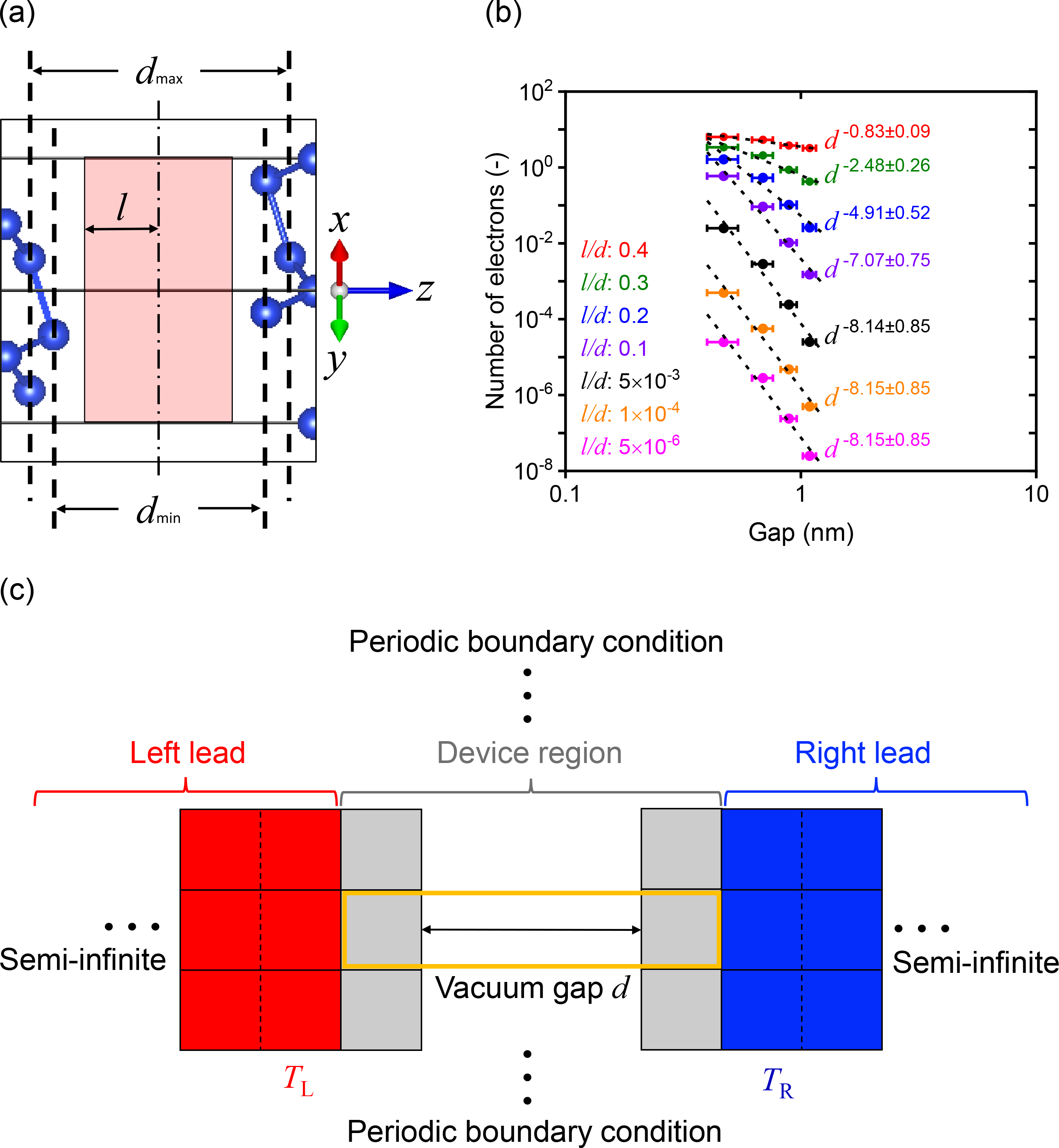}
\caption{(a) Vacuum gap region of a 1~${\times}$~1~${\times}$~8 supercell after structural optimization. The number of electrons is calculated within the volume specified by the red shaded box. The length $l$ surrounds the middle of the average vacuum gap. (b) Number of electrons as a function of the average vacuum gap thickness $d$ and the ratio $l/d$.  (c) The $zz$ components of the interatomic force constants acting on the atoms within the volume identified by the orange box are extracted, and their summations are reported in Fig. 3(d) of the main text.}
\label{Fig.S2}
\end{figure}

The number of electrons, $n_\mathrm{e}$, in the middle of the vacuum gap within the 1~${\times}$~1~${\times}$~8 supercell is calculated by integrating the electron density within the volume specified by the red shaded box in Fig.~S\ref{Fig.S2}(a). Specifically, the ratio $l/d$, where $l$ is defined in Fig.~S\ref{Fig.S2}(a) and $d$ is the average vacuum gap thickness, was varied from 0.4 to $5{\times}10^{-6}$. For $l/d$ values smaller than $5{\times}10^{-3}$, the number of electrons follows a $d^{-8.15 \pm 0.85}$ power law, as shown in Fig.~S\ref{Fig.S2}(b). This converging trend indicates that the number of electrons in the middle of the vacuum gap can be calculated using $l/d$ $<$ $5{\times}10^{-3}$. 

The interatomic force constants reported in Fig. 3(d) of the main text are calculated by extracting the $zz$ components of the interatomic force constants acting on the atoms contained in the volume of the device region specified by the orange box in Fig.~S\ref{Fig.S2}(c). Only the $zz$ components of the interatomic force constants are used since phonon transport across a single-digit nanometer vacuum gap is quasi-one-dimensional \cite{Chiloyan2015,Tokunaga2021}.

 \clearpage
\newpage

\section{S4. Phonon dispersion relations and phonon density of states of bulk silicon calculated via the density functional theory}
\begin{figure}[p!]
\centering
\vspace{100pt} 
\includegraphics[width=1\linewidth]{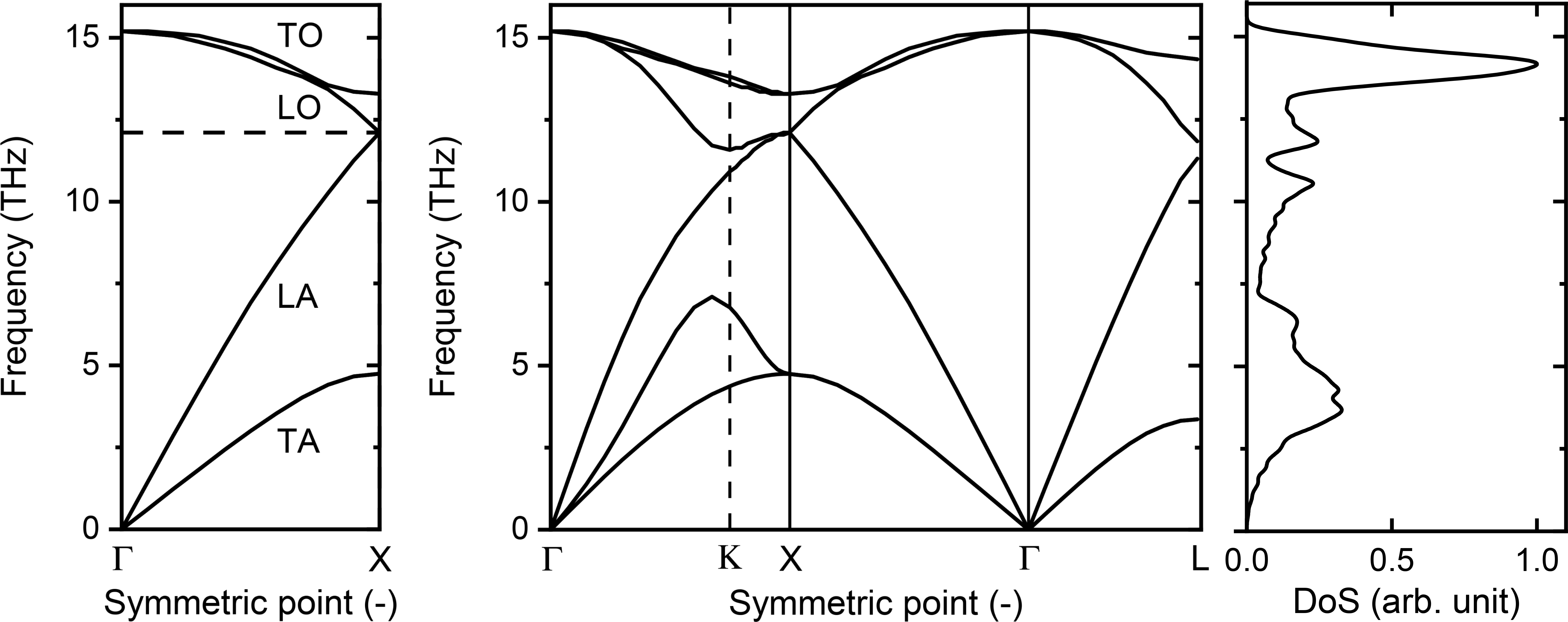}
\caption{Phonon dispersion relations and phonon density of states (DoS) of bulk Si obtained via the density functional theory \cite{Gonze2002,Gonze2020}. Transverse optical, longitudinal optical, longitudinal acoustic, and transverse acoustic phonon modes are respectively denoted as TO, LO, LA, and TA. The order of the symmetric points within the first Brillouin zone used for generating the phonon dispersion relations is selected based on Refs. \cite{Setyawan2010,Davis2011}. The horizontal dashed line is drawn at a frequency of 12 THz along the $\Gamma$-$\mathrm{X}$ direction.}
\label{Fig.S3}
\end{figure}

Figure S\ref{Fig.S3} shows the phonon dispersion relations and phonon density of states of bulk Si obtained via first-principles calculations based on the density functional theory. In this work, acoustic and optical phonon modes are characterized by frequencies respectively lower and higher than 12 THz. This threshold is defined based on the boundary between longitudinal acoustic (LA) and longitudinal optical (LO) phonon modes located around 12 THz along the $\Gamma$-$\mathrm{X}$ direction, as shown in Fig. S\ref{Fig.S3}.

 \clearpage 
\newpage

\section{S5. Phonon populations}
\begin{figure}[p!]
\centering
\includegraphics[width=0.9\linewidth]{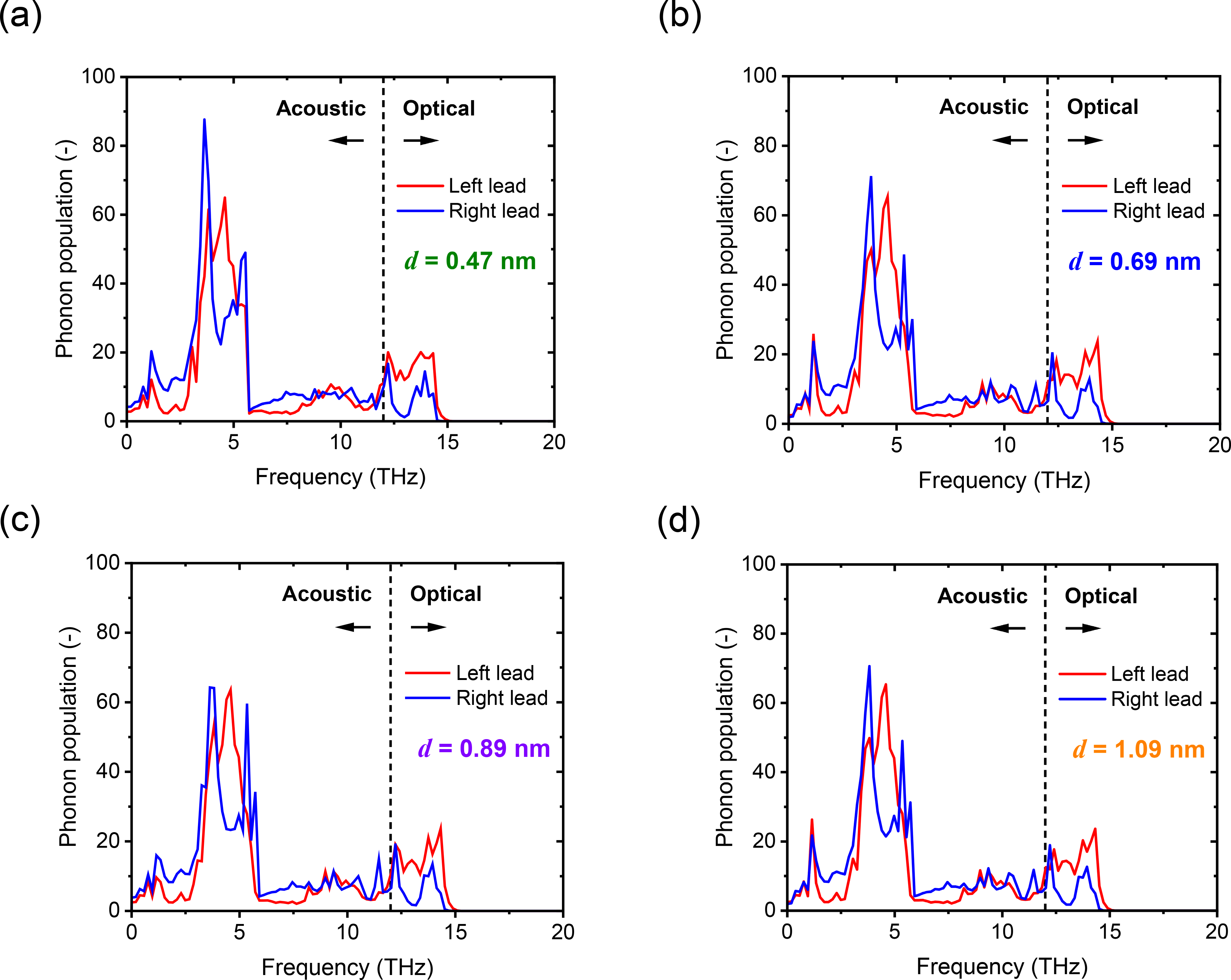}
\caption{Phonon populations of the atomic layers adjacent to the vacuum gap in the left and right leads for: (a) $d$ = 0.47 nm. (b) $d$ = 0.69 nm. (c) $d$ = 0.89 nm. (d) $d$ = 1.09 nm. The vertical dashed line at 12 THz delimits the frequencies associated with acoustic and optical phonons.}
\label{Fig.S4}
\end{figure}

Phonon populations of the atomic layers adjacent to the vacuum gap in the left and right leads are calculated for vacuum gaps of 0.47, 0.69, 0.89, and 1.09 nm and are reported in Fig.~S\ref{Fig.S4}. These atomic layers are denoted as layer 16 in the left lead and layer 1 in the right lead in Fig. 1(b) of the main text. Note that surface reconstruction is considered in the calculations. The phonon populations are obtained by multiplying the Bose-Einstein distribution function by the local phonon density of states of the device region obtained via the atomistic Green's function method. The local phonon density of states, $D_{\mathrm{d}}$, is calculated as follows \cite{Zhang_NHTPBF_2007}:
\begin{equation}
\label{pldos}
    D_{\mathrm{d}} = i\left(G_{\mathrm{d}}-G_{\mathrm{d}}^{\dagger}\right)\omega
\end{equation} where $G_{\mathrm{d}}$ is the Green's function of the device region, the superscript $\dagger$ denotes conjugate transpose, and $\omega$ is the phonon frequency. 

Figure~S\ref{Fig.S4} clearly shows that the populations of acoustic phonons characterized by frequencies lower than 12 THz largely surpass that of optical phonons for all vacuum gaps considered. 

 \clearpage
\newpage

\normalem 
\bibliography{SI}